\documentclass[letter]{aa}
\usepackage{CJK}
\newcommand{\gkai}[1]{\begin{CJK*}{UTF8}{gkai}\raisebox{.1em}{(}#1\raisebox{.1em}{)}\end{CJK*}}

\usepackage[colorlinks=true,linkcolor=blue,citecolor=blue, urlcolor=blue]{hyperref}
\usepackage{natbib,hyperref}

\usepackage{graphicx}
\usepackage{txfonts}
\defcitealias{2023andrae}{A23}
\defcitealias{2022matsuno}{M22}
\defcitealias{2018li}{L18}
\defcitealias{2023martin}{MS23}

\begin{document} 

%is your star anaemic? Don't come here to mine?
   \title{\textit{Gaia}'s brightest very metal-poor (VMP) stars}
   \titlerunning{\textit{Gaia}'s brightest very metal-poor stars}
   \subtitle{Metallicity catalogue of a thousand VMP stars from \textit{Gaia}’s radial velocity spectrometer spectra \thanks{Table \ref{table} released with this paper is only available in electronic form at the CDS via anonymous ftp to \href{https://cdsarc.cds.unistra.fr/}{cdsarc.u-strasbg.fr} (\href{ftp://130.79.128.5/}{130.79.128.5}) or via \href{http://cdsweb.u-strasbg.fr/cgi-bin/qcat?J/A+A/}{http://cdsweb.u-strasbg.fr/cgi-bin/qcat?J/A+A/}}}

   \author{Akshara Viswanathan
          \inst{1}
          \and
          Else Starkenburg\inst{1}\and
          Tadafumi Matsuno\inst{1}\and
          Kim A. Venn\inst{2}\and
          Nicolas F. Martin\inst{3}\inst{,}\inst{4}\and
          Nicolas Longeard\inst{5}\and
          Anke Ardern-Arentsen\inst{6}\and
          %Pascale Jablonka\inst{5,7}\and
          Raymond G. Carlberg\inst{7}\and
          Sébastien Fabbro\inst{8,2}\and
          Georges Kordopatis\inst{9}\and
          Martin Montelius\inst{1}\and
          Federico Sestito\inst{2}\and
          Zhen Yuan \gkai{袁珍}\inst{3}
          % \fnmsep\thanks{Just to show the usage
          % of the elements in the author field}
          }

   \institute{Kapteyn Astronomical Institute, University of Groningen, Landleven 12, 9747 AD Groningen, The Netherlands\\
              \email{astroakshara97@gmail.com}
         \and
             Dept. of Physics and Astronomy, University of Victoria, P.O. Box 3055, STN CSC, Victoria BC V8W 3P6, Canada\and
             Universit\'e de Strasbourg, CNRS, Observatoire astronomique de Strasbourg, UMR 7550, F-67000 Strasbourg, France \and
             Max-Planck-Institut f\"{u}r Astronomie, K\"{o}nigstuhl 17, D-69117 Heidelberg, Germany \and
             Laboratoire d'astrophysique, \'Ecole Polytechnique F\'ed\'erale de Lausanne (EPFL), Observatoire, 1290 Versoix, Switzerland\and
             Institute of Astronomy, University of Cambridge, Madingley Road, Cambridge CB3 0HA, UK\and
             %GEPI, Observatoire de Paris, Universit\'e PSL, CNRS, 5 Place Jules Janssen, 92195, Meudon, France\and
             Department of Astronomy \& Astrophysics, University of Toronto, Toronto, ON M5S 3H4, Canada\and
             National Research Council Canada, 5071 West Saanich Road, Victoria, Canada, V9E 2E7\and
             Universit\'e C\^ote d'Azur, Observatoire de la C\^ote d'Azur, CNRS, Laboratoire Lagrange, Nice, France
             % \thanks{The university of heaven temporarily does not
             %         accept e-mails}
             }

   \date{Received 11 September 2023; accepted 21 December 2023}

% \abstract{}{}{}{}{} 
% 5 {} token are mandatory
 
  \abstract
   {\textit{Gaia} DR3 has offered the scientific community a remarkable dataset of approximately one million spectra acquired with the radial velocity spectrometer (RVS) in the calcium II triplet region, which is well-suited to identify very metal-poor (VMP) stars. 
   However, over 40\% of these spectra have no released parameters by \textit{Gaia}'s GSP-Spec pipeline in the domain of VMP stars, whereas VMP stars are key tracers of early Galactic evolution.}
   {We aim to provide spectroscopic metallicities for VMP stars using \textit{Gaia} RVS spectra, thereby producing a catalogue of bright VMP stars distributed over the full sky that can serve as the basis for studies of early chemical evolution throughout the Galaxy.}
   {We selected VMP stars using photometric metallicities from the literature and analysed the \textit{Gaia} RVS spectra to infer spectroscopic metallicities for these stars.}
   {The inferred metallicities 
   %have a median measurement uncertainties of 0.05 dex and 
   agree very well with literature high-resolution metallicities with a median systematic offset of 0.1 dex and standard deviation of $\sim$0.15 dex. 
   The purity of this sample in the VMP regime is $\sim$80\%, with outliers representing a mere $\sim$3\%.}
   {We have built an all-sky catalogue of $\sim$1500 stars available, featuring reliable spectroscopic metallicities down to [Fe/H]$\sim$-4.0, of which $\sim$1000 are VMP stars. 
   More than 75\% of these stars have either no spectroscopic metallicity value in the literature to date or have been flagged as unreliable in their literature spectroscopic metallicity estimates. 
   This catalogue of bright (G<13) VMP stars 
   %increases the number 
   is three times larger than the current sample of well-studied VMP stars in the literature 
   %by more than a factor of three 
   in this magnitude range, making it ideal for high-resolution spectroscopic follow-ups and studies of the properties of VMP stars in different parts of our Galaxy.}


   \maketitle
%
%-------------------------------------------------------------------

\section{Introduction}

The study of very metal-poor stars (VMP, [Fe/H]<-2, ie., one-hundredth of solar metallicity) holds profound implications on several astrophysical processes as they provide a direct link to the early universe \citep[e.g.][]{2005beers}. They are thought to carry the imprint of the first supernovae \citep[see e.g. ][]{2018ishigaki}, provide us with relics of the smallest and earliest galaxies \citep[e.g.][]{2000chiba,2020yuan}, and offer essential indications regarding the characteristics of the first stars and their mass distribution \citep[see for a recent review][]{2023klessen}.
%These stars provide a direct link to the early universe as they offer insights into stellar evolution and the synthesis of heavy elements. 
%Born from elemental remnants of the first generations of stars, they can be interesting in many different ways: (i) they carry the imprint of the first supernovae \citep[see e.g. ][]{2018ishigaki}, (ii) they are relics from the era of the smallest, earliest galaxies that merged into the infant Milky Way \citep[e.g.][]{2000chiba,2020yuan,2023brauer}, (iii) they offer essential indications regarding the characteristics of the first stars and their mass distribution \citep[see for a recent review][]{2023klessen}. 
Bright VMP stars are of special interest to the community, because we can obtain spectra with high signal-to-noise ratios and subsequently unravel their chemical compositions in detail.
%, spatial distribution, and intrinsic properties using a sizeable sample of such VMP stars offers a unique pathway to the past of the universe. 
However, finding many of these bright, very and extremely metal-poor (VMP and EMP, [Fe/H]<-3) stars is challenging because they are rare among the more metal-rich and young populations of the Galaxy \citep{2004venn,2017youakim,2021yong,2021bonifacio}.

In recent decades, the Galactic archaeology community have striven to find more of these interesting and rare candidates, predominantly using the following three techniques: (i) mining large coverage spectroscopic surveys such as LAMOST \citep{2006zhao}, SDSS \citep{2000york}, RAVE \citep{2006steinmetz}, \textit{Gaia}-ESO survey \citep{2012gilmore}, APOGEE \citep{2017majewski}, and GALAH \citep{2015desilva}; (ii) prism or narrow-band photometric surveys looking at the metallicity sensitive calcium H and K lines region as in the HK Survey \citep{1992beers} and the Pristine survey \citep{2017starkenburg,2023martin} in the north and the Hamburg-ESO Survey \citep{2008christlieb} and SkyMapper survey \citep{2018wolf} in the south, along with more recent similar methods from S-PLUS \citep{2022almeida}, J-PLUS \citep{2019cenarro}, and J-PAS \citep{2014benitez} surveys, and (iii) using a mix of optical and infrared broad-bands to identify VMP candidates through their lack of molecular absorption near 4.6 microns \citep{2014schlaufman}.

As neither of these methods typically provide detailed high-resolution information sensitive to the most metal-poor regime, they are almost always combined with dedicated follow-up efforts \citep[e.g.][]{2019aguado, 2021yong, 2022li} to provide accurate metallicities and abundance patterns for stars pre-selected.

In this work, we combine photometric and spectroscopic efforts using some of the best datasets released by the \textit{Gaia} consortium in June 2022 for the purpose of finding and characterising VMP stars. 
The staggering \textit{Gaia} Data Release 3 \citep[DR3]{2023Gaia} released low-resolution spectra for about 220 million stars up to a magnitude of G$\sim$17.65 \citep{2022deangeli,2023aandrae} and medium-resolution spectra for about one million stars up to a magnitude of G$\sim$13 around the calcium triplet region \citep{2023recio}, making these spectra well-suited for the analysis of VMP stars down to metallicities of [Fe/H] $\sim$ -4.0 \citep[][]{2010starkenburg,2013carrera}. 
While the former dataset has been used to provide photometric metallicities \citep{2023andrae,2023aandrae,2023zhang,2023yao,2023martin}, the latter has been used by the \textit{Gaia} consortium to provide the largest (and the first space-based) dataset of stellar chemo-physical parameters \citep{2023recio}. 
We further leverage both catalogues in this work to analyse VMP stars specifically by pre-selecting them using photometric metallicities.
In doing so, we have ended up with about one thousand VMP stars confirmed by spectroscopic metallicities that are bright and spread over the entire sky. We have made this catalogue available to the community for high-resolution follow-ups and multiple science cases.

In Section \ref{2} we describe the method used to select and analyse these VMP stars. Section \ref{3} presents our results and a comparison of our metallicities with literature. Section \ref{4} summarises the properties of our VMP catalogue. 

\section{Methods}\label{2}

Approximately one million epoch-averaged RVS spectra have been released by the \textit{Gaia} consortium \citep{2023Gaia}, with the majority of them having undergone analysis and publication during \textit{Gaia} DR3. 
The analysis was performed using the General Stellar Parametriser for spectroscopy (GSP-Spec) module of the Astrophysical parameters inference system (Apsis) as described in \citet{2023recio}. 
However, a significant fraction of the spectra were not included in the DR3 publication of stars with GSP-Spec parameters. 
Here, we leverage the full DR3 sample of spectra with the aim to look for metal-poor stars. 
In the following subsections, we explore several approaches for identifying potential VMP targets. 
We accomplish this by utilizing photometric metallicities, selecting those with available \textit{Gaia} RVS spectra.
The main reason to use photometric pre-selection is to avoid reanalysing one million spectra, most of which are metal-rich \citep[see][for a similar approach using the LAMOST and Pristine surveys]{2023arentsen}. 
We then summarise the spectral analysis for the RVS spectra to infer metallicities.
\begin{figure}
\centering
\includegraphics[width=1.0\columnwidth]{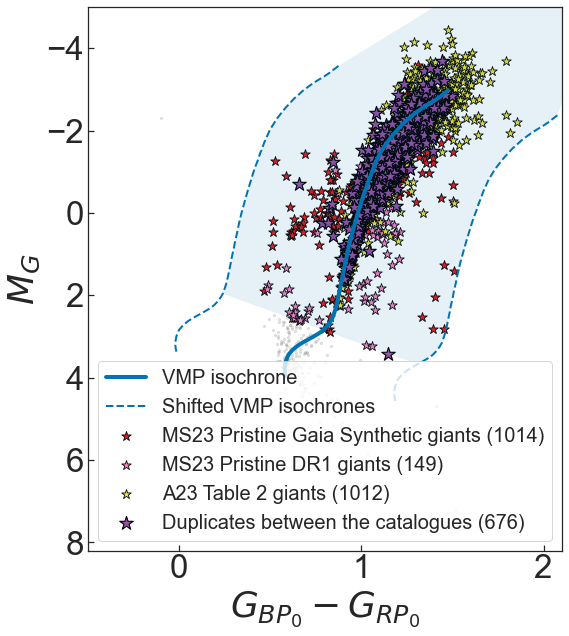}
\caption{Colour absolute magnitude diagram (CaMD) of stars that have photometric metallicities from \citetalias{2023martin} (Pristine \textit{Gaia} synthetic and Pristine DR1 catalogues) and \citetalias{2023andrae} Table 2 giants catalogue less than -2.0 dex. Duplicates between the three catalogues are shown as purple star symbols. The selection is justified by using  13 Gyr -2.2 [M/H] PARSEC isochrone shifted by $\pm0.6$ mag in $BP_0-RP_0$ and $\pm0.6$ mag in $M_G$. The blue polygon is the selection area for VMP stars analysed in this paper. The number of stars selected in each catalogue is shown in braces in their legend labels.}
          \label{selection}%
\end{figure}

\begin{figure*}
\centering
\includegraphics[width=1.0\textwidth]{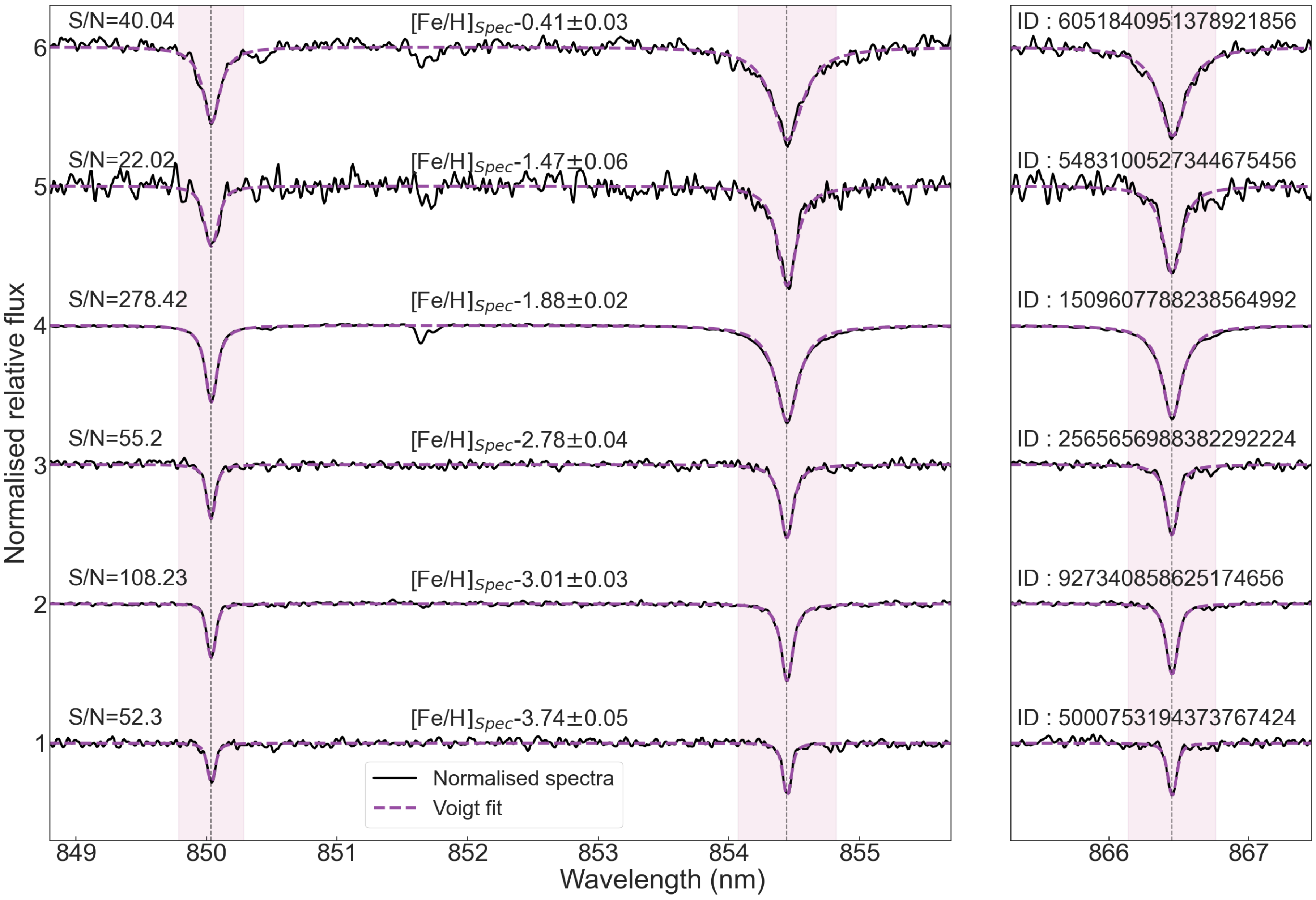}
\caption{\textit{Gaia} RVS spectra for six stars in this study. The stars are presented for a range of signal-to-noise ratio and sorted by inferred spectroscopic metallicities. For clarity, the normalised flux of each star is shifted by +1.0 and the spectra between the second and the third calcium triplet line is cut-off. The calcium triplet lines are highlighted in purple and central line is indicated by gray dashed lines.}
          \label{spectralfit}%
\end{figure*}

\subsection{Pre-selection of potential VMP stars}
To select VMP targets from the one million RVS spectra released, we use information from photometric metallicities inferred using published \textit{Gaia} XP (BP-RP) spectra using the XGBOOST algorithm by \citet[][hereafter, \citetalias{2023andrae}]{2023andrae} and Pristine survey model by \citet[][hereafter \citetalias{2023martin}]{2023martin}. 
In addition to this, we also use photometric metallicities using narrow band CaHK measurements in the northern hemisphere from the Pristine Survey data release 1 (DR1) released by \citetalias{2023martin}. 
We chose these three catalogues for our pre-selection due to their reliability in metallicity measurements down to the VMP regime for red giants \citep{2019aguado,2020venn,2021kielty,2022lucchesi,2023martin}.

\subsubsection{Photometric metallicities using the XGBoost algorithm}
\citetalias{2023andrae} published data-driven estimates of photometric metallicities for approximately 175 million stars using low-resolution XP spectra from \textit{Gaia} DR3. 
They train an XGBoost algorithm on APOGEE stellar parameters, supplemented with VMP stars from \citet{2022li}, utilizing various data features including spectral coefficients, narrowband fluxes, broadband magnitudes, CatWISE magnitudes, and parallax. 

We select star with photometric metallicities less than -2.0 (\texttt{mh\_xgboost<-2.0}) from Table 2 of \citetalias{2023andrae}, which consists of 17 million bright red giants (G<16) with precise and pure metallicity measurements.  
We perform extinction correction for these stars using the method described and performed on a reduced proper motion halo catalogue by \citet{2023viswanathan}. 
The extinction correction involves calculating the "extinction fraction" based on the dust density model, scaling the \citet{1998schlegel} 2D dust maps, and applying an extinction curve. 
These corrections allow us to calculate the amount of foreground dust for each star as a function of its parallax and location on the sky. 
Extinction corrected magnitudes are necessary to infer metallicities as discussed later in this section.

\subsubsection{Photometric metallicities using synthetic CaHK ran through the Pristine survey model}
Using \textit{Gaia} DR3 XP spectra data, \citetalias{2023martin} calculated synthetic CaHK magnitudes for approximately 219 million stars. 
These synthetic magnitudes combined with broadband \textit{Gaia} information are pushed through the Pristine survey model to yield photometric metallicities. 
We applied the following recommended cuts on the Pristine \textit{Gaia} synthetic catalogue released by \citetalias{2023martin}:
%dfeh<0.3&Pvar<0.3&abs(Cstar)<Cstar_1sigma&mcfrac>0.8&ebv<0.5&ruwe<1.4& RPlx>5&FeHphot_\textit{Gaia}<-2
\begin{itemize}
    \item Photometric metallicity [Fe/H] less than -2.0 dex (\texttt{FeHphot\_CaHKsyn}<-2.0 dex)
    \item Fraction of Monte Carlo iterations used to determine [Fe/H] uncertainties is greater than 0.8 (\texttt{mcfrac\_CaHKsyn}>0.8)
    \item   Photometric metallicity uncertainty less than 0.3 dex (0.5*(\texttt{FeH\_CaHKsyn\_84th} - \texttt{FeH\_CaHKsyn\_16th})<0.3 dex)
    \item       Probability of being a variable star being less than 30\% \texttt{Pvar}<0.3
    \item Extinction on B-V magnitude is less than 0.5 (\texttt{E(B-V)}<0.5)
    \item Photometric quality cut that is defined as {$C^*$}<{$\sigma_{C^*}$} (abs(\texttt{Cstar})<\texttt{Cstar\_1sigma})
\end{itemize}

\subsubsection{Photometric metallicities using the Pristine survey DR1}
The Pristine data release 1 (DR1) comes with metallicities calculated using Pristine CaHK narrow band and \textit{Gaia} broad band magnitudes for all the \textit{Gaia} stars with released XP spectra within the Pristine survey footprint.
We use the same cuts as for Pristine \textit{Gaia} synthetic catalogue for the Pristine DR1 catalogue to select potential VMP stars.
The Pristine DR1 selection adds few stars that are VMP candidates and have low quality synthetic CaHK magnitudes. 

\subsubsection{Final selection}
For all three photometric metallicity catalogues, we use a \textit{Gaia} astrometric quality cut \texttt{RUWE}<1.4 and parallax cut \texttt{parallax\_over\_error>5}. 
Hot targets and dwarfs are removed using shifted VMP PARSEC \footnote{\href{http://stev.oapd.inaf.it/cgi-bin/cmd}{stev.oapd.inaf.it/cgi-bin/cmd}} isochrones \citep{2017marigo} as shown in Figure \ref{selection}. 
We keep the isochrone selection as wide as possible to pick up as many VMP candidate stars as possible at the risk of picking up metal-rich contaminants. 
The final sample of stars to be analysed using the published \textit{Gaia} RVS spectra are plotted as yellow, red and pink star symbols for the respective photometric catalogues they belong to. 
In the main VMP red giant branch, we see a large overlap between the different catalogues as expected. These duplicates are shown in purple.
This comparison with all three catalogues of photometric metallicities and analysis of various stricter cuts on the Pristine catalogues and their effect on the CaMD is summarised later in Appendix \ref{b}. We note that most of the relatively metal-rich outliers (that have inferred spectroscopic metallicities that are much larger than the photometric estimates) in the sample are outliers in the Colour-absolute Magnitude Diagram (CaMD). Our final sample has 1014 stars from the \citepalias{2023andrae} catalogue, 1012 stars and 149 stars from the Pristine \textit{Gaia} synthetic and Pristine DR1 catalogues from \citepalias{2023martin} respectively. 
About 676 of them exist in at least two of the three catalogues.

\subsection{Spectral analysis pipeline}
The \textit{Gaia} consortium released RVS spectra normalised for continuum. 
We renormalised the spectra using a spline representation as described in Appendix \ref{a} to avoid systematic offsets seen when validated with high-resolution metallicities. 
We ended up with 1441 stars that are potential VMP candidates for which we analyse the \textit{Gaia} RVS spectra to infer metallicities.

For the next step, we utilized a modified pipeline based on \cite{2022longeard}  %(see Viswanathan et al., (in prep) for more details on the method) 
to fit equivalent widths to all three calcium triplet lines, namely, 850.04 nm, 854.44 nm and 866.45 nm at the same time and also computed additional radial velocity offsets in the process. 
Initially, we created smoothed spectra by applying a Gaussian kernel and focus on the calcium triplet (CaT) lines. 
These lines are modeled using Voigt profiles and their positions were determined by minimizing the difference between a simulated spectrum containing only the CaT lines and the observed spectrum. 
The initial radial velocity estimate was obtained through cross-correlation and is typically close to zero as the publicly available \textit{Gaia} RVS spectra are already in the rest frame. 
However owing to the handful of stars with radial velocity offsets greater than 5 km/s, we also released the radial velocity offset and error on this parameter as a part of this catalogue. 
%This estimate serves as a starting point for further analysis and improving the accuracy of the radial velocity measurement. 
A Markov chain Monte Carlo (MCMC) algorithm is then employed to fit the observed spectra, with the aim of deriving the radial velocity offset, depth, and full width at half maximum, and, eventually, the equivalent widths for the three calcium triplet lines. 
Constraints are applied to ensure the relative depths and widths of the lines are consistent such as the depth of the first line should be smaller than the second and the third which is in turn smaller than the second. 
The MCMC analysis was performed for each star, and the best-fit values are determined based on the maximum likelihood. 

The equivalent widths (EWs) of the CaT lines are converted into metallicity measurements using the calibration provided by \cite{2013carrera}.
This calibration for metallicities from the CaT equivalent width is an empirical relation based on observations of 55 metal-poor field stars in high resolution where the Fe I and Fe II spectral lines (which are weaker lines that are less affected by NLTE than the Ca II triplet lines) are measured using a resolution R>20,000. Therefore, the metallicities inferred from this calibration will be more close to an NLTE equivalent. 
Thus, this calibration works very well for VMP red giant stars and, due to its empirical nature, is equivalent to NLTE analyses. 
It requires magnitudes, calcium triplet equivalent widths and distances (inverted parallax) or height above or below the horizontal branch as inputs.
The magnitudes are corrected for extinction following the procedure described in \cite{2023viswanathan}, which is also summarised in detail in the previous subsection.
To go from \textit{Gaia} G magnitude to Johnston-Cross V or I magnitudes, we use the conversion defined by \citealt{2021riello}.
Uncertainties associated with the metallicity measurements are determined through a Monte Carlo procedure that takes into account the uncertainties in the equivalent widths, photometry, colour, distance (i.e. parallax uncertainties), and a calibration relation -- which are the input parameters in converting EWs to metallicities. 
The resulting probability distribution function (PDF) captures the uncertainty in the metallicity determination, with the standard deviation (using a Gaussian approximation) representing the uncertainty on the metallicities.
Figure \ref{spectralfit} illustrates some typical spectra obtained and published by \textit{Gaia} RVS after renormalisation as a black line together with the best fit Voigt profile by the pipeline described above as a purple dashed line. 
This subsample is chosen based on a mix of signal-to-noise ratios (S/Ns) and inferred metallicities of our sample.
The median S/N ratio of our sample is 36.2, while the lowest and highest are 15.3 and 278.4, respectively. 
About 70\% of the stars have a S/N lower than 50, which makes our MCMC fitting method robust on these spectra.

\section{Validation}\label{3}
%In this section, we compare our inferred spectroscopic metallicities from the \textit{Gaia} RVS spectra with other \textit{Gaia} RVS spectra analysis in literature. 
%We examine the number of relatively metal-rich outliers using comparisons with spectroscopic surveys such as APOGEE DR17 \citep{2017majewski,2019wilson,2022abdurro}, and GALAH DR3 \citep{2021buder}. 
%To study the precision of our metallicities in the very metal-poor regime, we use high-resolution spectroscopic follow-up of VMP stars by the SkyMapper survey \citep{2021yong}, the R-process Alliance (RPA) survey \citep{2018hansen,2018sakari} and the LAMOST survey VMP stars pipeline \citep{2018li} using the Subaru telescope \citep{2022li}, in addition to the SAGA database \citep{2008suda} of very metal-poor sources. \footnote{\href{http://sagadatabase.jp/}{sagadatabase.jp}} 

\begin{figure}
\centering
\includegraphics[width=1.0\columnwidth]{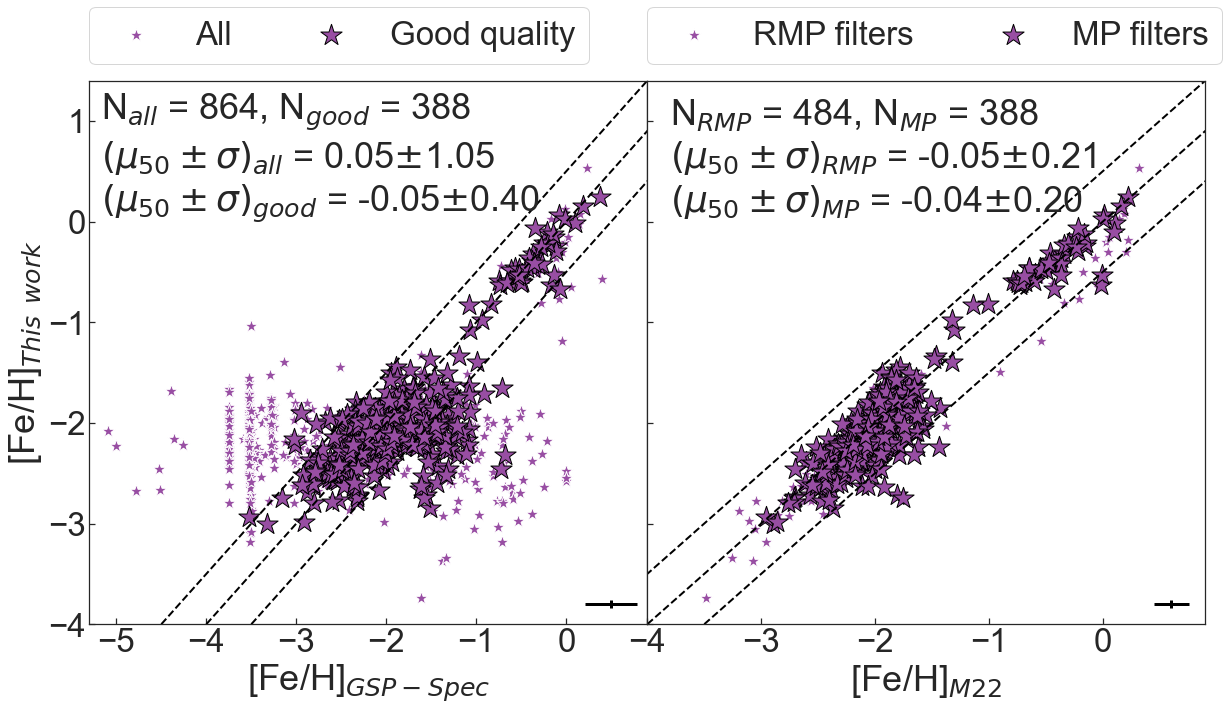}
\caption{Comparison of our metallicities with other \textit{Gaia} based spectroscopic metallicities such as GSP-Spec pipeline (left) and improved metallicity estimates for VMP stars in GSP-Spec catalogue by \citetalias{2022matsuno} (right). The number of stars (with stricter quality cuts), median ($\mu_{50}$), and 1$\sigma$ standard deviation ($\sigma$) in $\Delta$[Fe/H] and median uncertainties on the metallicity measurements (where available) is indicated at the top and bottom of each panel. The dashed lines shows the 1:1 line and corresponding $\pm$0.5 dex offsets.}
          \label{tm-comp}%
\end{figure}

\subsection{Comparison with other \textit{Gaia} based spectroscopic metallicity catalogues}
%The \textit{Gaia} consortium analysed the spectra taken with RVS instrument using the pipeline, General Stellar Parametriser-spectroscopy. 
About 71.20\% of the entire \textit{Gaia} RVS spectra, 5.6 million stars, have been analysed by the GSP-Spec module. 
This percentage decreases as the metallicity decreases (down to 40\% in the VMP end) and the \textit{Gaia} RVS spectra are only made publicly available for a small subset of well-behaved objects (about 12.70\% of the entire GSP-Spec objects).  
We do not find any significant trend between the fraction of un-analysed stars, or stars with bad solutions, and the S/N of the \textit{Gaia} RVS spectra.
About 78\% of the 5.6 million GSP-spec analysed stars have the most reliable metallicities after a very strict filtering (see Fig. 26 in \citealt{2023recio}). 
Due to the restricted wavelength range and lack of spectral information, metallicity estimates for VMP stars analyzed by the GSP-Spec module suffer from parameter degeneracy and exhibit large measurement uncertainties and systematic offsets \citep{2011kordopatis}. 
Additionally, the recommended quality cuts filter out a significant portion of these stars due to confusion with hot stars or challenges posed by cool K and M-type giants. 

To address this, \citet[][hereafter \citetalias{2022matsuno}]{2022matsuno} aimed to break the parameter degeneracy from lack of spectral information by incorporating photometric and astrometric information and reanalyzing FGK-type stars in the GSP-Spec catalog. This approach resulted in more precise metallicity estimates, reducing uncertainties and improving agreement with high-resolution literature values. 
The inclusion of photometric information proved valuable in overcoming the challenges posed by lack of spectral information for VMP stars.

Because the GSP-Spec module and the catalogue from \citetalias{2022matsuno} use the same spectra (directly or indirectly) to obtain metallicity and atmospheric parameters, we compare those metallicities to our inferred metallicities. 
Figure \ref{tm-comp} shows a comparison of our inferred metallicities with GSP-Spec metallicities and \citetalias{2022matsuno} metallicities. In the left panel, the large symbols stand for stars that pass the recommended quality cuts by \citealt{2023recio} and small star symbols represent all the stars with published GSP-Spec parameters. In the right panel, the large symbols stand for stars that pass the recommended GSP-Spec quality cuts as mentioned previously \citepalias[defined as MP filters by][]{2022matsuno} and the small symbols are the stars with relaxed criteria improved by adding photometric information \citepalias[defined as RMP filters by][]{2022matsuno}. 
Both catalogues have relatively small median metallicity offsets for stars that pass the quality cuts recommended. Meanwhile the GSP-Spec catalogue has higher dispersion compared to the metallicities from \citetalias{2022matsuno}. The sample size in overlap between these catalogues is small ($\sim$26\%). This is expected due to the parameter degeneracy from lack of spectral information and thus higher uncertainties in the inferred metallicities at the VMP end. 

\begin{figure}
\centering
\includegraphics[width=1.0\columnwidth]{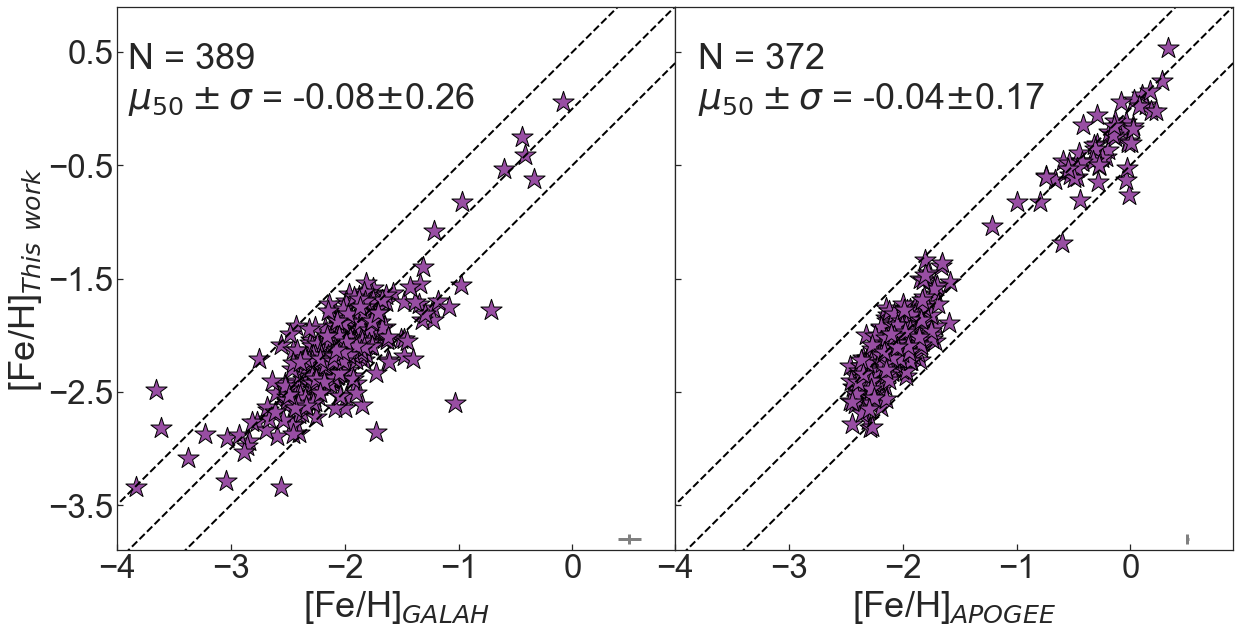}
\includegraphics[width=1.0\columnwidth]{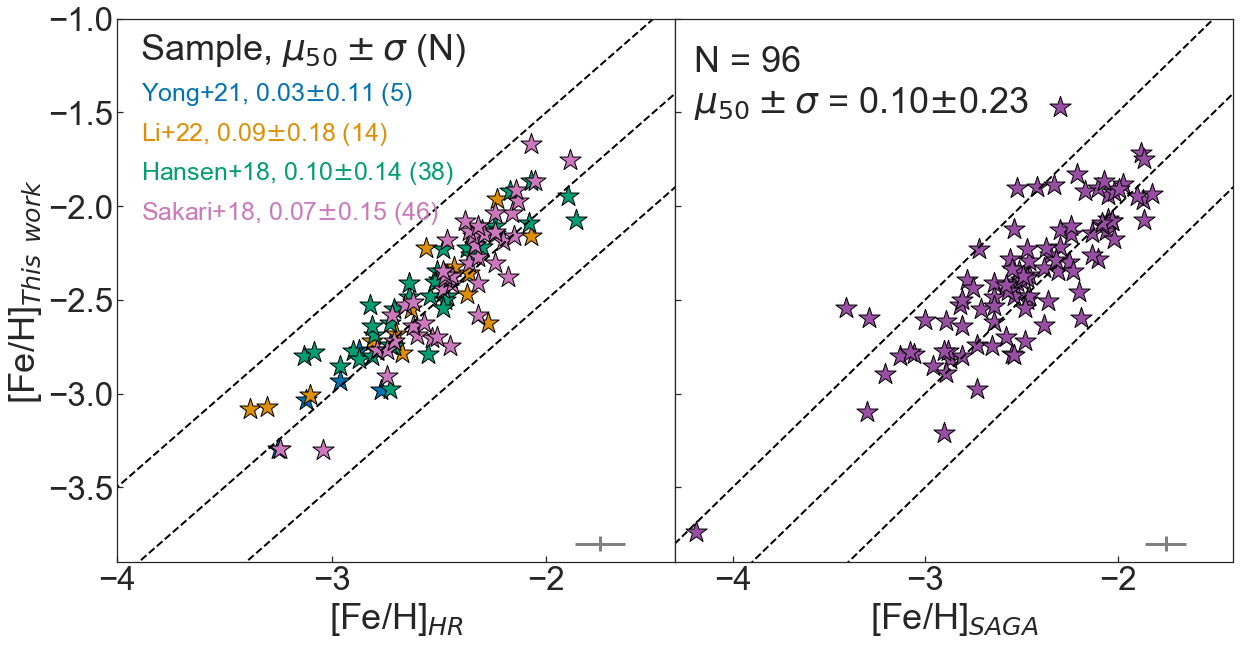}
\caption{Validation of our metallicities with existing spectroscopic surveys such as GALAH DR3 (top left), and APOGEE DR17 (top right), and high-resolution spectroscopic samples of metal-poor stars from \cite{2018hansen, 2018sakari,2021yong,2022li} (bottom left), and SAGA database of VMP stars (bottom right). The number of stars, median ($\mu_{50}$) and 1$\sigma$ standard deviation ($\sigma$) in $\Delta$[Fe/H] and median uncertainties on the metallicity measurements (where available) is indicated at the top and bottom of each panel. The dashed lines shows the 1:1 line and corresponding $\pm$0.5 dex offsets.}
          \label{hr-comp}%
\end{figure}

\subsection{Comparison with spectroscopic surveys and high-resolution VMP catalogues}
Next, we validated our metallicities and examine the number of relatively metal-rich outliers using comparisons with spectroscopic surveys such as APOGEE DR17 \citep{2017majewski,2019wilson,2022abdurro}, and GALAH DR3 \citep{2021buder}.
We removed stars with \texttt{STAR\_BAD} or \texttt{FE\_H\_FLAG} flagged from the APOGEE sample and those with \texttt{flag\_sp} $\neq$ 0 or \texttt{flag\_fe\_h} $\neq$ 0 from the GALAH sample. 
The top panels of Figure \ref{hr-comp} shows a comparison of our inferred metallicities with the existing spectroscopic surveys along with the coressponding median offset ($\mu_{50}$) and one-sigma standard deviation ($\sigma$, using a  Gaussian approximation) at the top of each panel. 
We can see the comparison with GALAH DR3 in the top left and APOGEE DR17 in the top right panels. 
The comparison with APOGEE stops at -2.5 dex which is up to where the ASPCAP pipeline assigns metallicities. 
In the comparison with GALAH, we see a few outliers especially in the EMP end which could be due to largely featureless GALAH spectra for EMP stars and because the GALAH DR3 pipeline is not tailored towards EMP stars with no/weak metal lines \citep{2022hughes}. 
However, we visually inspected the fits and MCMC chains for these stars, and are confident in the metallicities we assign for them. 
The comparison with both APOGEE and GALAH shows the robustness of our metallicities in all metallicity regimes, including few metal-rich stars that we picked up as outliers in the photometric selection.
The agreement with APOGEE (median offset of -0.04) is better than with GALAH (median offset of -0.08). 
Part of this offset might also be due to NLTE versus LTE analyses (our method provides metallicities close to the NLTE analyses).
Nevertheless, there are no catastrophic outliers given the width of the distributions ($\sigma$$\sim$0.2).
The comparison with these existing spectroscopic surveys show that our inferred metallicities are very reliable especially in the VMP regime.
Less than 3\% of the stars in low metallicity regime ([Fe/H]$_{This}$ $_{work}$<-1.5) are outliers ([Fe/H]$_{GALAH}$ or [Fe/H]$_{APOGEE}$>-1.5). 

The precision in our metallicity determinations is examined by comparison with results in four homogeneously analysed high-resolution spectroscopic follow-up catalogues from \citet{2021yong}, \citet{2022li}, \citet{2018hansen}, \citet{2018sakari}.
Comparisons with these four catalogues allow us to study the precision of our metallicities at the VMP end and also to quantify the systematic error in our method because these are large sets of homogeneously analysed stars, as opposed to assembling individual follow-up from various sources in the literature.
From this comparison (see Figure \ref{hr-comp} bottom left panel), we can see that our metallicities are accurate down to the EMP regime with median offsets as low as 0.07 dex with the largest crossmatch that we have with the \citet{2018sakari} catalogue of metal-poor stars. 
This offset might also be due to NLTE versus LTE analyses as mentioned previously.
We also compare our metallicities with the SAGA database of metal-poor stars in the literature. 
It is noteworthy that three of our four catalogues from the previous comparison are already a part of the SAGA database -- except the \citet{2022li} catalogue. 
However, this helps us quantify the precision of our metallicities collectively with a large number of stars from the literature. 
The comparison with the SAGA database shows that our metallicities are robust in the VMP regime and are ready to be used for an all-sky study of metal-poor stars and ideal for high-resolution spectroscopic follow-up given their relative brightness.
From these comparisons, we recommend using 0.1 dex as systematic uncertainties in our metallicities. 
A similar comparison from the point of view of metallicity difference and how much the offset has improved over the GSP-Spec released metallicities is presented in Appendix \ref{c}.
%-0.26+0.41-gspspec-461+57
%-0.07+0.20-rvs-662(409)(660)+214(179)(189)-57(42)

\begin{figure}
\centering
\includegraphics[width=1.0\columnwidth]{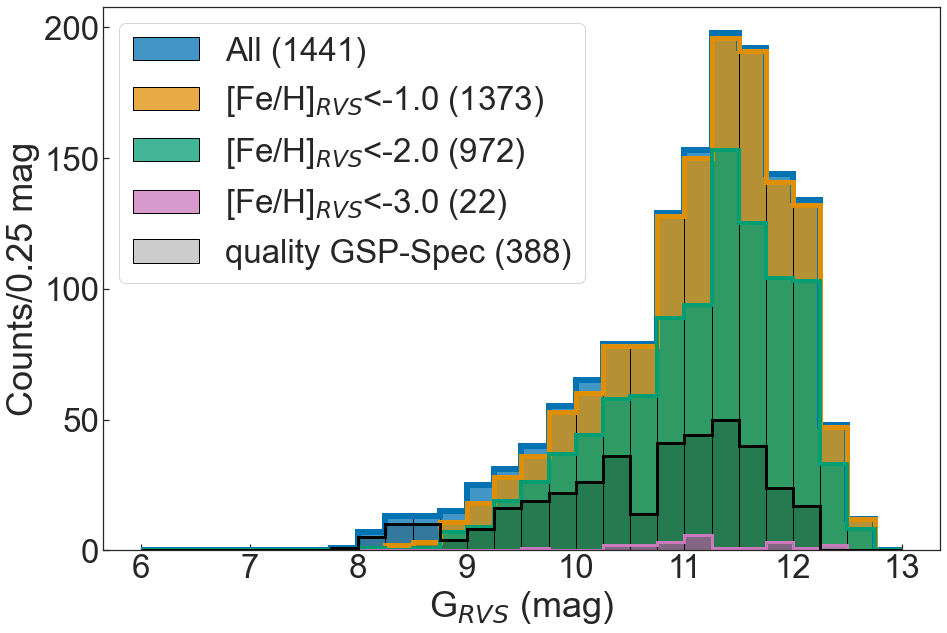}
\includegraphics[width=1.0\columnwidth]{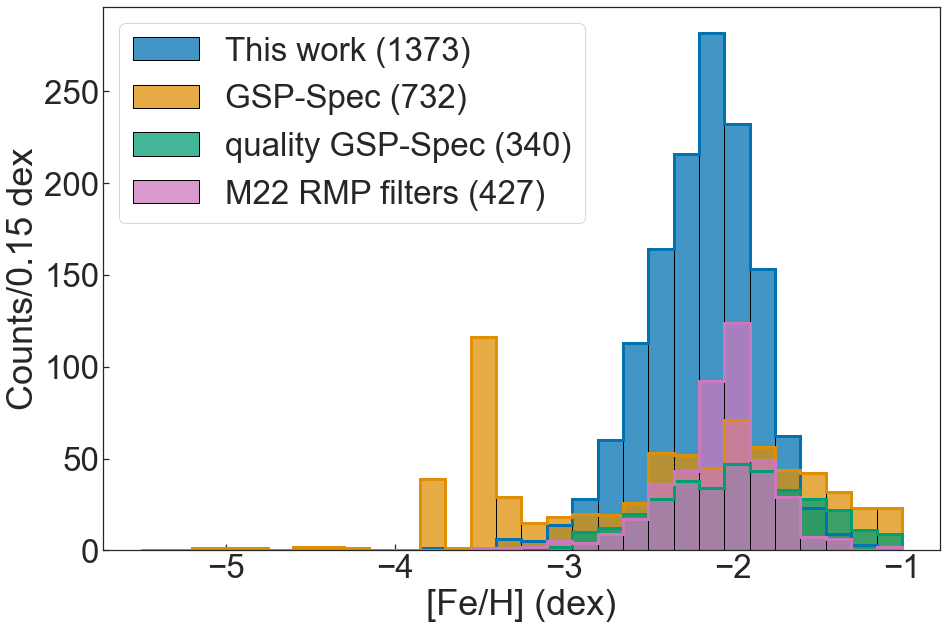}
\caption{Magnitude distribution of metal-poor stars in our VMP catalogue into groups of MP, VMP and EMP stars and magnitude distribution of stars with reliable metallicities from published GSP-Spec information (top). Metallicity distribution of our VMP catalogue and subsample with GSP-Spec published stars, good quality GSP-Spec stars, and reanalysed \citetalias{2022matsuno} VMP stars (bottom). Note: the metallicity distributions consists of metallicities inferred by each of the catalogues mentioned in the legend.}
          \label{gmag-feh}%
\end{figure}

%hansensakariliyong catalogue: hansen:1-38,sakari:

\section{Discussion and conclusions}\label{4}

%In this section, we summarise the improvement in precision and number over the existing literature values and discuss the properties of our catalogue made publicly available here. We summarise our conclusions and future prospects in the end. 

\subsection{Properties of the \textit{Gaia} RVS VMPs catalogue}

The VMP catalogue presented here has a brightness range of 6th to 13th magnitudes in the G band. 
Our catalogue of VMP stars is three times bigger than the number of VMP stars known in the literature in this brightness range (based on a comparison with the SAGA database).
The median measurement uncertainty in spectroscopic metallicities inferred in our catalogue is $\sim$0.05 dex.
The top panel of Fig. \ref{gmag-feh} shows the distribution of G$_{RVS}$ magnitudes for the whole catalogue and different subsets such as for metal-poor (MP, [Fe/H]<-1), VMP and EMP stars. 
We also show the distribution of G magnitudes for a subset of stars that have a released and reliable GSP-Spec analysis.
From this distribution, we can see that, typically, the released, but previously un-analysed, RVS spectra go up to a magnitude fainter (G$\sim$13) than available GSP-Spec analysed stars (up to G$\sim$12). Over 40\% of the stars analysed in this work have no published GSP-Spec analysis and more than 75\% of them do not pass the quality cuts recommended by \citet{2023recio} (as described for their Figure 26) for the usage of the parameters published by GSP-Spec analysis.
The re-analysis of \citetalias{2022matsuno} of GSP-Spec results adds more reliable metallicities in the VMP regime by breaking the temperature degeneracy from lack of spectral information with external (non-\textit{Gaia}) data and reduces this fraction to about 65\%. This re-analysis evidently performs better at the VMP end. However, because 80\% of their stars do not have a released RVS spectra, they cannot visually check their spectral fits which might make it unreliable in some special cases. This is not a limitation in our method. The bottom panel of Figure \ref{gmag-feh}, clearly demonstrates the large numbers of VMP stars with reliable metallicities that are added to the available datasets from GSP-Spec analysis (along with the quality cuts) and the re-analysis done by \citetalias{2022matsuno}. It is important to note that the clump of EMP stars from GSP-Spec analysis are unreliable and disappear in the histogram with recommended quality cuts. %This catalogue is  for the homogeneous study of metal-poor stars in different parts of the Galaxy and for high-resolution follow-up of VMP stars to study the other chemical abundances that help understand the history and origin of some of the oldest stars in the Galaxy. 
 
%This shows the power of \textit{Gaia} RVS spectra which will be made available public in the upcoming \textit{Gaia} data releases. 
%We find about 1000 VMP stars and about 20 EMP stars in this work. 

% \section{Conclusion and Outlook}\label{5}
\subsection{Summary and outlook}
With the recent \textit{Gaia} DR3, the \textit{Gaia} consortium released about one million spectra obtained by the radial velocity spectrometer (RVS) instrument. In this paper, we use these publicly available \textit{Gaia} RVS spectra, with the main objective to provide an all-sky catalogue of bright VMP stars. To select potential VMP stars, we use publicly available photometric metallicities catalogues.

We present reliable metallicities for 1374 MP ([Fe/H]<-1.0), 973 VMP ([Fe/H]<-2.0), and 22 EMP ([Fe/H]<-3.0) stars. We recommend to add to the reported measurement uncertainty (only $\sim$0.05 in the median), a systematic uncertainty of 0.1 dex derived from comparison with high-resolution analyses. This is one of the largest and only all-sky catalogue of homogenously analysed VMP stars using spectroscopy and, for the first time, using \textit{Gaia} RVS spectra for a dedicated analysis of VMP stars. Our bright ($6<G_{RVS}<13$) VMP stars catalogue increases the number of known VMP stars in this brightness range by more than a factor of three when compared to the SAGA database and is homogeneously analysed for stars over the whole sky. In our sample, over 75\% of our stars has no reliable and/or no spectroscopic metallicities in the literature, 40\% have no available spectroscopic parameters at all, and 93\% of them have no high-resolution chemical abundances available in SAGA.

This work shows the potential of utilizing publicly accessible (archival) spectra to investigate the Galaxy's most metal-poor stars. Our catalogue is ideal for high-resolution spectroscopic follow-up due to its brightness range (meaning these stars require lower exposure times to get several other chemical abundances: few VMP stars are bright enough to be seen with a naked eye) and to study the all-sky distribution of metal-poor stars and their origin. As forthcoming \textit{Gaia} data releases will unfold in the coming years with many more spectra (about a
factor ten larger in DR4), we anticipate delving even deeper and substantially expanding our understanding on the origin of these very metal-poor stars in impressive numbers. %It is a golden age to do Galactic Archaeology. %Through this study, we navigate the tapestry of time, gaining insights into the evolution and growth of the Galaxy. \ES{(nice sentence, but we don't do any of this in this paper ;)}

%\section{Data availability} This catalogue is made available in this temporary repository: \texttt{\href{https://astroakshara.github.io/rvs-paper/\textit{Gaia}-RVS-VMP-catalogue-AV23b-vSep23.csv}{astroakshara.github.io/rvs-paper/\textit{Gaia}-RVS-VMP-\\catalogue-AV23b-vSep23.csv}} in the format shown in Table \ref{table} before the acceptance of this paper.

\begin{table*}
\caption{Description of the columns of the \textit{Gaia} RVS spectra VMP stars catalogue made available publicly in this work} 
%(currently accessible in a temporary
%repository before acceptance \href{https://astroakshara.github.io/rvs-paper/\textit{Gaia}-RVS-VMP-catalogue-AV23b-vSep23.csv}{astroakshara.github.io/rvs-paper/\textit{Gaia}-RVS-VMP-catalogue-AV23b-vSep23.csv}). \\}
\label{table}
\centering
\begin{tabular}{llll}
\hline \hline
 Column & Description & Unit & Type \\
 \hline
source\_id & \textit{Gaia} DR3 Source ID & NA & longint \\
ra & \textit{Gaia} DR3 right ascension in ICRS (J2016) format & degrees & float \\
dec & \textit{Gaia} DR3 declination in ICRS (J2016) format & degrees & float \\
G\_0 & de-reddened \textit{Gaia} G magnitude & unitless & float\\
BP\_0 & de-reddened \textit{Gaia} G$_{BP}$ magnitude & unitless & float\\
RP\_0 & de-reddened \textit{Gaia} G$_{RP}$ magnitude & unitless & float\\
parallax & \textit{Gaia} DR3 parallax & mas & float\\
parallax\_error & Uncertainty on the \textit{Gaia} DR3 parallax & mas & float\\
snr & Signal-to-noise ratio of the renormalised \textit{Gaia} RVS spectra & unitless & float \\
v\_offset & Radial velocity offset from the rest frame & km/s & float \\
dv\_offset & Uncertainty in the radial velocity offset from the rest frame & km/s & float \\
ew1 & Equivalent width of the first calcium triplet line around 850.035 nm & nm & float \\
dew1 & Uncertainty on the equivalent width of the first calcium triplet line around 850.035 nm & nm & float \\
ew2 & Equivalent width of the second calcium triplet line around 854.444 nm & nm & float \\
dew2 & Uncertainty on the equivalent width of the second calcium triplet line around 854.444 nm & nm & float \\
ew3 & Equivalent width of the third calcium triplet line around 866.452 nm & nm & float \\
dew3 & Uncertainty on the equivalent width of the third calcium triplet line around 866.452 nm & nm & float \\
feh & Spectroscopic metallicity derived in this work using \textit{Gaia} RVS spectra & unitless & float \\
dfeh & Measurement uncertainty associated with the spectroscopic metallicity derived & unitless & float \\
\hline
\end{tabular}
\end{table*}

\begin{acknowledgements}
      We express our gratitude to the reviewer for dedicating their valuable time and providing insightful contributions that significantly enhanced the quality of our manuscript. AV thanks Ewoud Wempe for helpful discussions. ES acknowledges funding through VIDI grant "Pushing Galactic Archaeology to its limits" (with project number VI.Vidi.193.093) which is funded by the Dutch Research Council (NWO). 
      NFM and ZY gratefully acknowledge support from the French National Research Agency (ANR) funded project ``Pristine'' (ANR-18-CE31-0017) along with funding from the European Research Council (ERC) under the European Unions Horizon 2020 research and innovation programme (grant agreement No. 834148). AA acknowledges support from the Herchel Smith Fellowship at the University of Cambridge and a Fitzwilliam College research fellowship supported by the Isaac Newton Trust.
      This research was supported by the International Space Science Institute (ISSI) in Bern, through ISSI International Team project 540 (The Early Milky Way).
      This work has made use of data from the European Space Agency (ESA) mission \textit{Gaia} (https://www.cosmos.esa.int/\textit{Gaia}), processed by the \textit{Gaia} Data Processing and Analysis Consortium (DPAC, https://www.cosmos.esa.int/web/\textit{Gaia}/dpac/consortium). Funding for the DPAC has been provided by national institutions, in particular the institutions participating in the \textit{Gaia} Multilateral Agreement.
      AV thanks the availability of the following packages and tools that made this work possible: \texttt{vaex} \citep{2018vaex}, \texttt{astropy} \citep{2022astropy}, \texttt{NumPy} \citep{2006numpy,2011numpy}, \texttt{SciPy} \citep{2001scipy}, \texttt{matplotlib} \citep{2007matplotlib}, \texttt{seaborn} \citep{2016seaborn}, \texttt{JupyterLab} \citep{2016jupyter}, and \texttt{topcat} \citep{2018topcat}.
      
      % \textbf{Need to change referencing style to A\&A - it's MNRAS right now but I like the hyperlinks}
\end{acknowledgements}

\bibliographystyle{aa}
\bibliography{aanda}

\begin{appendix}

\section{Renormalisation of \textit{Gaia} RVS spectra}\label{a}
\begin{figure*}
\centering
\includegraphics[width=1.0\textwidth]{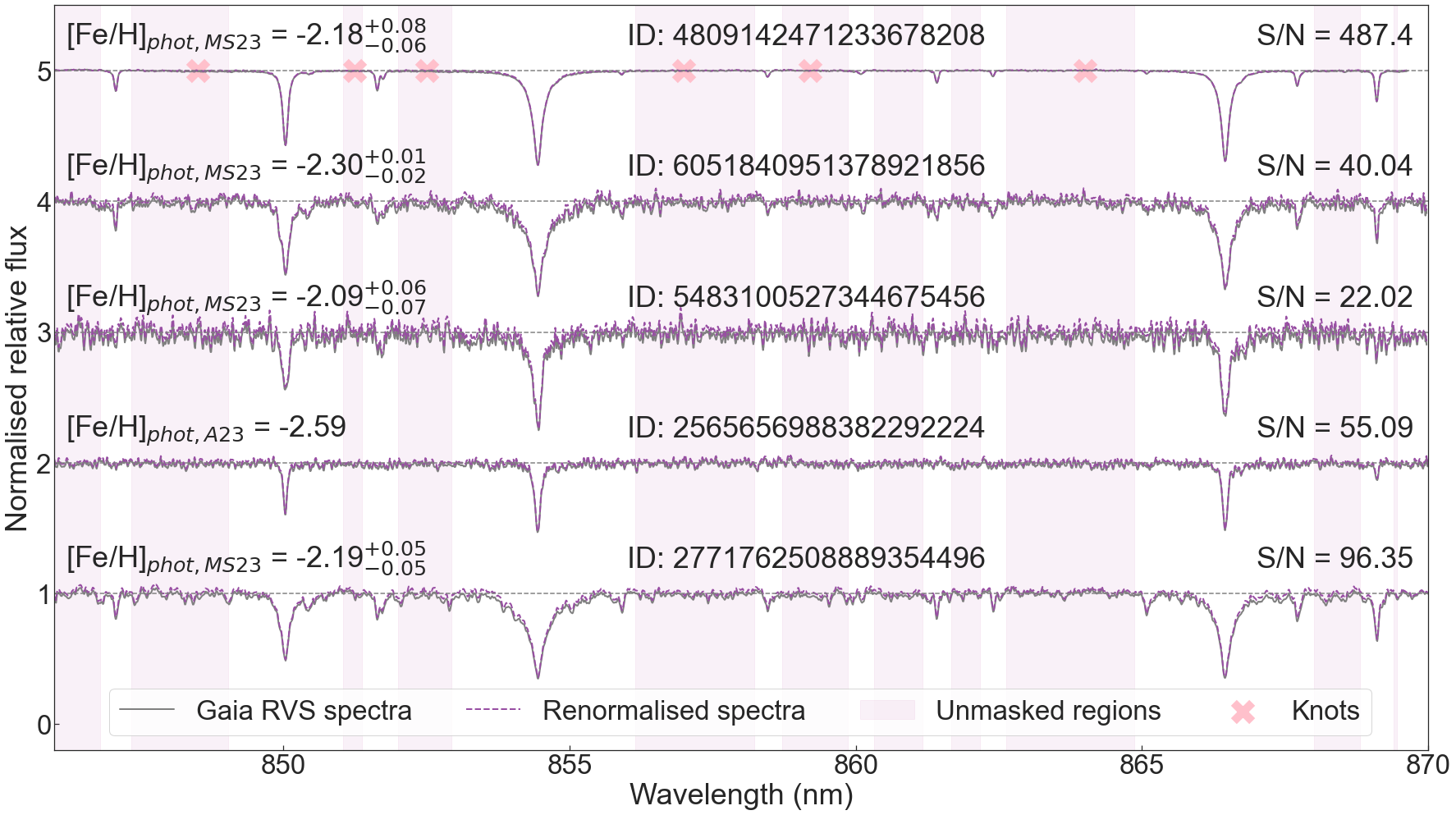}
\caption{\textit{Gaia} RVS spectra in gray of a selected subset of five stars chosen with varying signal-to-noise and metallicities. Renormalised \textit{Gaia} spectra in purple. The topmost spectra is the highest signal-to-noise spectra used to choose knots and masks to perform spline interpolation to infer the continuum. Gray lines represent the released \textit{Gaia} RVS spectra and purple lines represent the renormalised RVS spectra using spline representation. Purple bands show the unmasked region to define the continuum and red cross symbols show the chosen knots.}
          \label{normalisation}%
\end{figure*}

%spl = splrep(xx,yy,task=-1,t=knots)
%continuum = splev(xx, spl)
The \textit{Gaia} consortium released RVS spectra normalised for continuum. 
As we have the advantage of focusing on metal-poor stars, generally with less absorption lines and a more clear defined continuum, we re-normalised the spectra for our own purposes as follows. 
After masking the major absorption lines, $\pm$1 nm, $\pm$1.5 nm, $\pm$1.25 nm around the three calcium triplet lines, respectively, and $\pm$0.2 nm around other lines, we fit a spline representation of the spectrum using the \texttt{splrep} function in python's \texttt{scipy} module to a very high signal-to-noise spectrum.
Using this spectrum, we define the following knots covering the continuum regions for the spline representation: [848.50, 851.25, 852.50, 857.00, 859.20, 864.00] nm. 
The same knots and fitting technique are applied to all spectra for re-normalisation. 
This type of normalisation has very few assumptions and works well for stars of different S/N and different metallicities in the metal-poor end as can be seen in Figure \ref{normalisation}.
The knots and masked regions, as well as some released and re-normalised spectra are shown in Figure \ref{normalisation}. 
We note that tests of the obtained metallicity results compared to high-resolution datasets (such as those performed in Section \ref{3}) perform better after the re-normalisation. 
In particular, a systematic offset of $\sim$0.25 dex is alleviated significantly and almost fully disappears. 
We choose to leave out 4 stars from our parent sample that have a signal-to-noise ratio higher than 300, because we believe there to be a risk of over-correcting the re-normalisation, especially if they are not very metal-poor. 
%It may be possible that there is a risk of over-correcting by renormalisation.
%However, we use this spline interpolation method because it has the least assumptions to begin with, given that \textit{Gaia} has not published the impact of the first normalization on metallicity estimation. 

\section{Comparison with input photometric metallicities}\label{b}

\begin{figure*}
\centering
\includegraphics[width=1.0\textwidth]{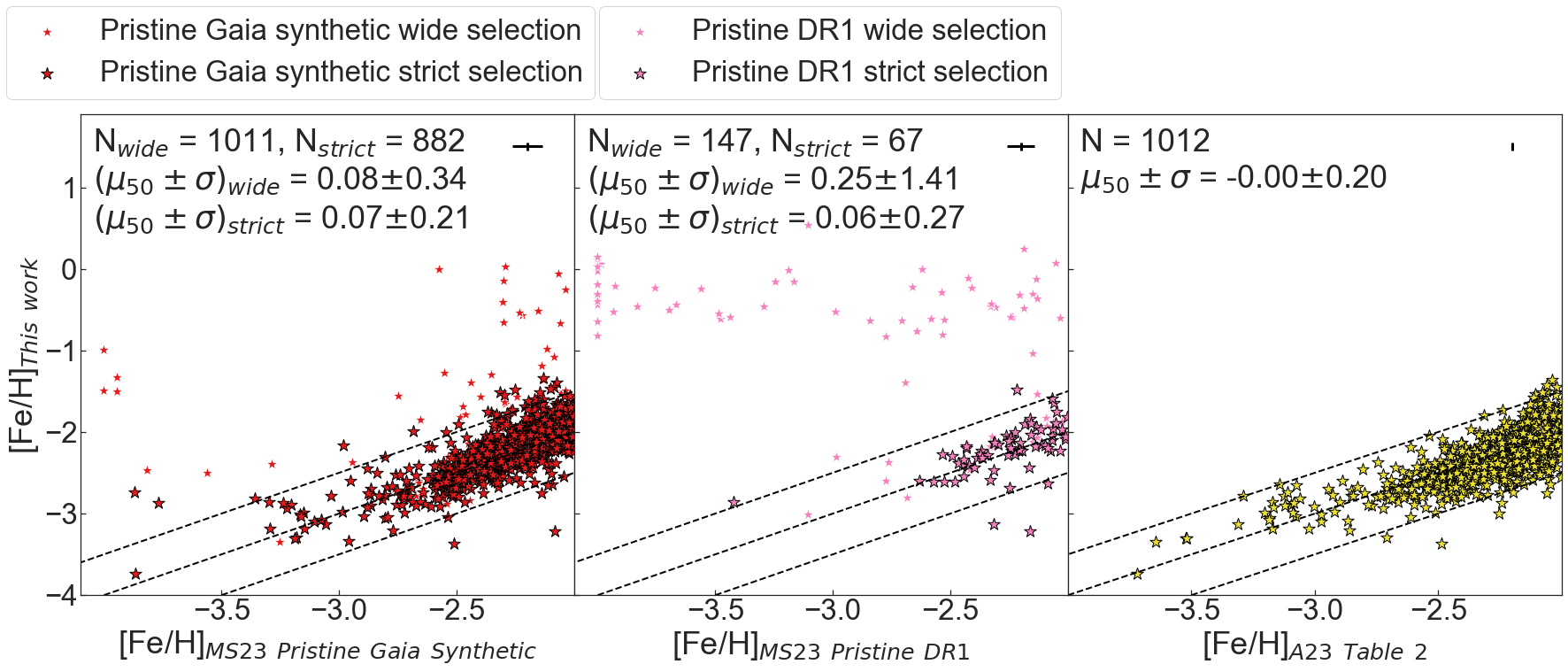}
\caption{Comparison of our metallicities with photometric metallicities from \textit{Gaia} XP based catalogues such as Pristine \textit{Gaia} Synthetic catalogue from \citetalias{2023martin} (left), Table 2 giants catalogue from \citetalias{2023andrae} (right) and dedicated CaHK based metallicity survey such as Pristine DR1 from \citetalias{2023martin} (middle). The number of stars, median ($\mu_{50}$) and 1$\sigma$ standard deviation ($\sigma$) in $\Delta$[Fe/H] and median uncertainties on the metallicity measurements (where available) is indicated at the top of each panel. The dashed lines shows the 1:1 line and coressponding $\pm$0.5 dex offsets.} 
          \label{phot-comp}%
\end{figure*}

\begin{figure}
\centering
\includegraphics[width=1.0\columnwidth]{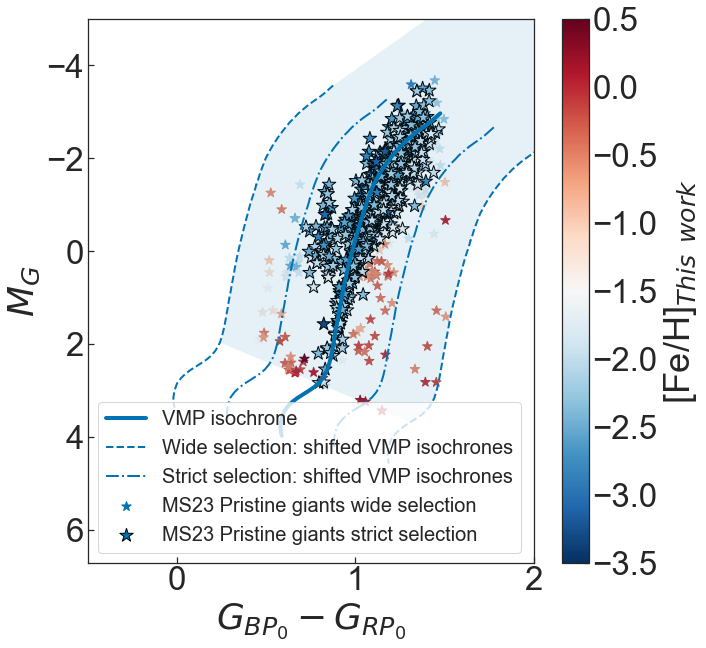}
\caption{CaMD diagram of stars that have photometric metallicities from \citetalias{2023martin} (Pristine \textit{Gaia} synthetic and Pristine DR1 catalogues) less than -2.0 dex, showing the consequence of wide and strict selections, colour-coded by the derived spectroscopic metallicities.}
          \label{isochrone-appendix}%
\end{figure}

Figure \ref{phot-comp} shows the comparison of our metallicities with photometric metallicities from Pristine \textit{Gaia} synthetic catalogue, Pristine DR1 catalogue from \citetalias{2023martin} and XGBOOST catalogue from \citetalias{2023andrae} on the left, middle and right panels respectively. 
We can also see the number of stars in each catalogue, median offset and one sigma deviation at the top of each panel.

While the sample from \citetalias{2023andrae} we use here has already very strict quality cuts applied, the \citetalias{2023martin} selection is initially kept quite broad in order to recover as many VMP stars as possible. This is reflected in a larger number of catastrophic failures from these catalogues. 
However, if we apply stricter quality cuts on variability (\texttt{Pvar}<0.2), Monte Carlo iteration within the colour-colour space (\texttt{mcfrac}=1.0), error on observational or synthetic CaHK magnitudes (\texttt{d\_CaHK}<0.3), CASU photometric data reduction flag (\texttt{flag}=-1, denoting very likely point-sources - only for Pristine DR1), and, moreover, if we remove stars that have an assigned metallicity equal to -4.0 dex (assigned to stars that lie outside but close to the theoretical no metals line), it can be seen that the performance significantly improves without sacrificing too many good VMP and EMP stars. 
If we additionally tighten our initial red giant branch selection (illustrated in Figure \ref{isochrone-appendix}), the performance is even better. 
This can be visualised in Figure \ref{phot-comp} as small and large symbols for wide and strict quality cuts on the Pristine catalogues (left and middle panels). 
We can already see from the plots that the strict cut reduces the median offset and standard deviation on both the Pristine catalogues. 
From Figure \ref{isochrone-appendix}, we can see that almost all the catastrophic outliers in the Pristine catalogues are outliers in the CaMD as well. 
Stricter shifted VMP isochrone selection justify this approach. 

%For the Pristine DR1 photometric metallicities, we use an additional CASU flag (\texttt{flag}=-1) which is a flag that the CASU pipeline assign for the morphology of the source: flag=-1 for very likely stars, -2 for likely stars, or -9 for saturated stars with extrapolated photometry, +1 for a source that is likely extended. 
%We didn't use this flag in our initial selection to see if we can recover some VMP stars from saturated sources.
%It is important to note that our initial photometric selection is kept as wide as possible as the goal is recover as much VMP stars as possible and not to make a pure photometric selection.
%These additional cuts make the offset and dispersion between our metallicities and photometric metallicities from \citetalias{2023martin} catalogue much lower. 

For Pristine DR1 photometric metallicities, the median offset reduces from 0.25 dex to 0.06 dex and the deviation reduces from 1.40 dex to 0.27 dex after these stricter quality cuts and the largest contribution is made by the stricter selection on the CASU flag. 
%This difference is mostly attributed to the additional CASU flag selection.
For the Pristine \textit{Gaia} synthetic catalogue, the deviation reduces from 0.34 dex to 0.24 dex after stricter quality cuts.
%Almost all the outliers lie outside the main VMP red giant branch.\textcolor{red}{Maybe a figure in the appendix to show this?}
These extra cuts bring the performance of the \citetalias{2023martin} catalogues at a similar level to the more strict selection of the XGBOOST catalogue by \citetalias{2023andrae}, as can be appreciated from a comparison of the three panels in Figure \ref{phot-comp}.

\section{Comparison with other photometric metallicities}\label{d}

\begin{figure}
\centering
\includegraphics[width=\columnwidth]{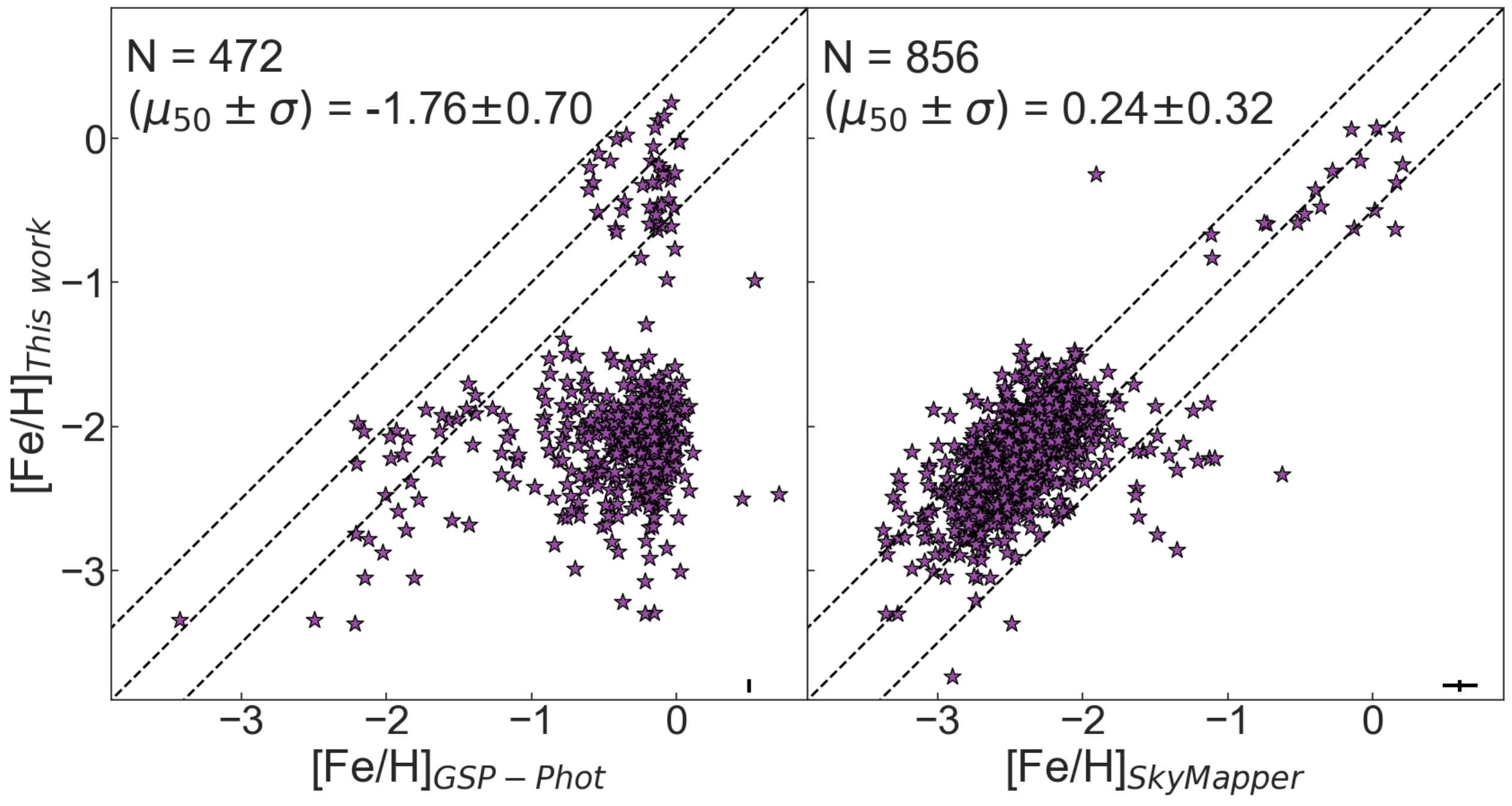}
\caption{Comparison of our metallicities with photometric metallicities from \textit{Gaia} XP based GSP-Phot pipeline \citep{2023aandrae} and SkyMapper Southern Survey \citep[SMSS,][]{2022huang}. The number of stars, median ($\mu_{50}$) and 1$\sigma$ standard deviation ($\sigma$) in $\Delta$[Fe/H] and median uncertainties on the metallicity measurements (where available) is indicated at the top of each panel. The dashed lines shows the 1:1 line and coressponding $\pm$0.5 dex offsets.} 
          \label{other-phot}%
\end{figure}

Figure \ref{other-phot} shows the comparison of our metallicities with photometric metallicities from the GSP-Phot pipeline based on \textit{Gaia} XP spectra\citep{2023aandrae} on the left and SkyMapper Southern Survey \citep[SMSS,][]{2022huang}. The number of stars, median offset, and one sigma standard deviations is also indicated at the top of each panel.

A comprehensive collection of metallicities for \textit{Gaia} sources with XP spectra was first introduced in DR3 \citep{2023aandrae}. The [M/H] values were derived using synthetic model spectra in comparison with the XP spectra, aiming for a consistent approach to stellar parameter estimates across the color-magnitude diagram (CMD). However, subsequent external validation has revealed significant limitations in these [M/H] values, including systematics and a notable prevalence of "catastrophic" outliers. Two factors acknowledged by the authors contributing to these shortcomings were (i) despite detailed knowledge, the \textit{Gaia} XP system exhibits imperfections, resulting in substantial discrepancies between the predictions of synthetic models and XP data, consequently leading to inaccurate [M/H] estimates and (ii) due to varying information content about [M/H] at low resolution, and especially for certain temperatures (e.g. OB stars), XP spectra may lack informativeness on [M/H] for specific Teff and log g combinations. This is also reflected in our comparison with almost 88\% of the stars being outliers (differing more than 0.5 dex in value). The photometric-metallicity estimates from \citet{2022huang} for approximately 24 million stars with over 19 million dwarf stars and 5 million giant stars are derived from stellar colors sourced from SMSS DR2 and \textit{Gaia} EDR3, utilizing training datasets with spectroscopic metallicity measurements from high-resolution surveys (PASTEL \citep{2016soubiran} and SAGA \citep{2008suda}) as well as low or medium-resolution spectroscopic surveys (LAMOST and SDSS/SEGUE). Through extensive external testing, their typical uncertainty range between 0.20 and 0.25 dex for VMP giant stars. This is reflected in the comparison with our spectroscopic metallicities with median offset and one sigma deviation down to 0.24 and 0.32 dex, respectively.

\section{Precision of this work over \textit{Gaia} GSP-Spec pipeline for VMP stars}\label{c}

\begin{figure}
\centering
\includegraphics[width=1.0\columnwidth]{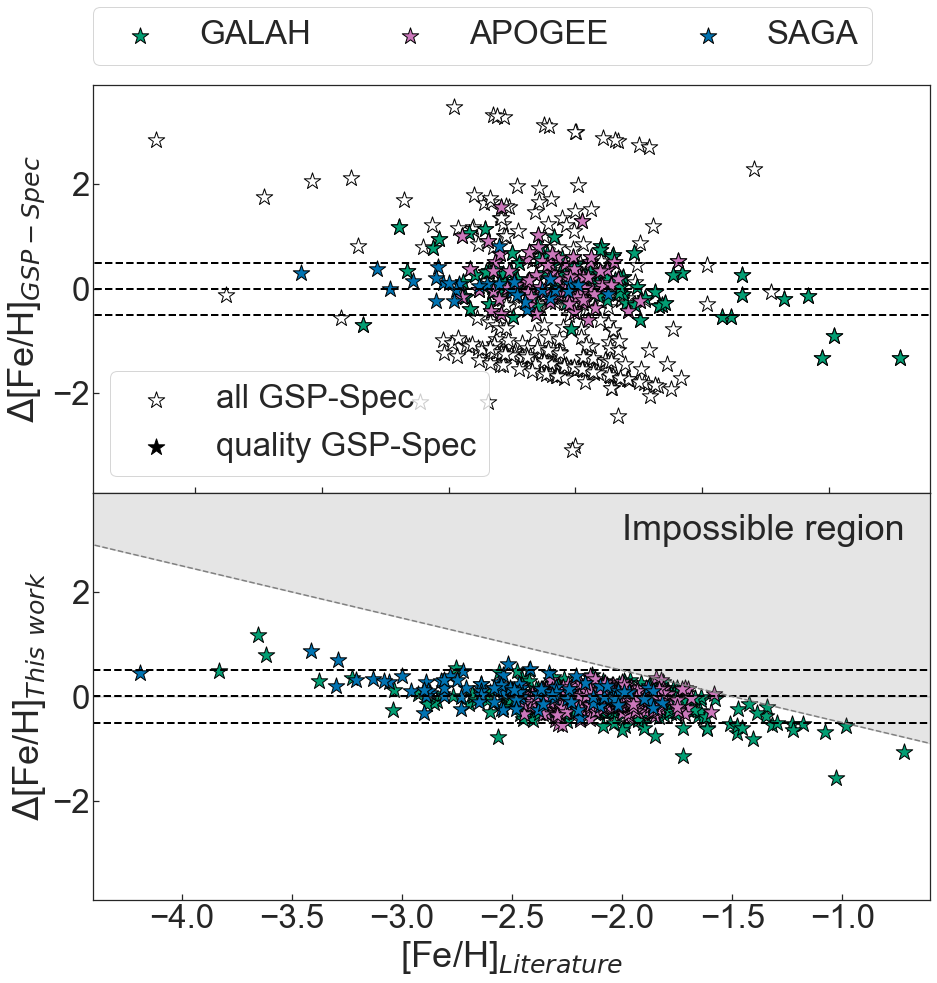}
\caption{Precision of our metallicities (bottom) over GSP-Spec metallicities (top) validated using literature metallicities. The dashed lines shows the 0.0$\pm$0.5 dex metallicity difference lines. The gray shaded region in the bottom panel indicates the part of the figure where it's impossible to have data points due to the low metallicity selection ([Fe/H]$_{This}$ $_{work}$<-1.5).}
          \label{gspspecvsrvs}%
\end{figure}

In Figure \ref{gspspecvsrvs}, we show the metallicity difference $\Delta$[Fe/H] \footnote{Throughout this paper, $\Delta$[Fe/H] is always this work minus other work} (our metallicity minus the metallicity from literature) as a function of literature metallicities (from GALAH DR3, APOGEE DR17, and SAGA database) for both the GSP-Spec metallicities and metallicities derived using published RVS spectra in this work.
Note: this figure only includes stars that have an inferred metallicity from this work less than -1.5 ([Fe/H]$_{This}$ $_{work}$<-1.5) to look at the precision of our metallicities in the VMP regime. 
However, this introduces a selection function in the gray shaded region in the bottom panel of Figure \ref{gspspecvsrvs}. 
This also explains the scatter with GALAH metallicities at [Fe/H]$_{Literature}$$\sim$-1.0. 
The scatter with GALAH metallicities in the EMP end is mostly due to lack of spectral information in GALAH at these lowest metallicities.
From this comparison, we can see the precision of metallicities in our catalogue over the GSP-Spec metallicities especially for stars that do not pass the quality cuts described in \citet{2023recio} which makes up about half of the catalogue in the VMP end. This is also reflected in the large uncertainties provided by these metallicities (large errors on these stars are mostly caused by a lack of spectral information).
 %First appendix

% In this case....
% \begin{figure*}
% \centering
% \includegraphics[width=16.4cm,clip]{1787f24.ps}
% \caption{Plotted above...}
% \label{appfig}
% \end{figure*}

% Because the optical images...
%Second appendix

% These studies, however, have faced...

% The second method produces...
\end{appendix}
\end{document}